\documentstyle[psfig]{article}
\begin{document}
\begin{center}
{\bf GRB 000301c with peculiar afterglow emission } \\

{\sc R. Sagar$^1$, V. Mohan$^1$, S. B. Pandey$^1$, A. K. Pandey$^1$, C. S. 
Stalin$^1$ and A. J. Castro-Tirado$^2$}

\end {center}

\centerline {\it $^1$U. P. State Observatory, Manora Peak, Nainital -- 263 129,
 India}
\centerline {\it $^2$IAA-CSIC, P.O. Box 03004, E-18080, Granada, Spain}
\bigskip

\begin{abstract}
The CCD magnitudes in Johnson $V$ and Cousins $R$ and $I$ photometric passbands 
are determined for GRB 000301C afterglow starting $\sim$ 1.5 day after the 
$\gamma-$ray burst. In fact we provide the earliest optical observations for 
this burst. Light curves of the afterglow emissions in $U, B, V, R, I, J $ and 
$K^{'}$ passbands are obtained by combining the present measurements with the 
published data. Flux decay shows a very uncommon variation relative to other 
well observed GRBs. Overall, there is a steepening of the optical and 
near-infrared flux decay caused by a geometric and sideways expanding jet. This 
is superimposed by a short term variability especially during early time 
($\Delta t < 8$ days). The cause of variability is not well understood, though
it has occurred simultaneously with similar amplitude in all the filters. After 
removing the superposed flux due to variability, we derive the early and late 
time flux decay constants using jet model. The late time flux decay is the 
steepest amongst the GRB OTs observed so far with $\alpha \sim 3$. Steepening 
in the flux decay seems to have started simultaneously around $\Delta t \sim 
7.6$ day in all passbands. On the other hand no such variations are observed in 
the quasi-simultaneous broad-band photometric spectral energy distributions of 
the afterglow. The value of spectral index in the optical-near IR region is 
$\sim -1.0$. Redshift determination with $z=2.0335$ indicates cosmological 
origin of the GRB having a luminosity distance of 16.6 Gpc. Thus it becomes the 
second farthest amongst the GRBs with known distances. An indirect estimate of 
the fluence $>$ 20 keV indicates, if isotropic, $\ge 10^{53}$ ergs of release 
of energy. The enormous amount of released energy will be reduced, if the 
radiation is beamed which is the case for this event. Using a jet break time of 
7.6 days, we infer a jet opening angle of $\sim$ 0.15 radian. This means the
energy released is reduced by a factor of $\sim$ 90 relative
to the isotropic value. 
\end{abstract}

{Keywords: Photometry -- GRB afterglow -- flux decay -- spectral index }

\section {Introduction}

Gamma-ray bursts (GRBs) are short and intense flashes of cosmic high energy 
($\sim$ 10 KeV$-$10 GeV) photons. The study of GRBs was revolutionized in 1997 
when the Italian-Dutch $X-$ray satellite BeppoSAX started providing positions 
of some GRBs with an accuracy of a few arcminutes within a few hours after the 
burst. This made the multi-wavelength observations of the long-lived emission, 
known as afterglow of GRB, at longer wavelengths as a routine. Such 
observations are of crucial importance for understanding and constraining the 
active emission mechanisms of GRBs as well as for the study of the nature, 
structure and composition of surroundings. For this, early light curves
of GRB afterglows need to be densely sampled for long time intervals. A huge
amount of observing time is therefore required on optical telescopes. Since
the optical transient (OT) of a GRB has generally apparent $R$ magnitude
between 18 to 22, if it is detected within a day or so after the burst, the 1-m
class optical telescopes equipped with modern CCD detector are capable of
observing them. Such telescopes are not only large in number throughout the
world but also less in demand compared to 2-m class or larger size optical 
telescopes. The large amount of observing time is therefore available on
them (cf. Sagar 2000 for detailed discussions). The 1-m class optical
telescopes equipped with  CCD detector, thus, can contribute to the world-class
science in an emerging front-line research area of GRB. We at U.P. State 
Observatory (UPSO), Nainital, therefore, started the optical follow-up 
observations of the GRB afterglows in January 1999 under an international 
collaborative programme coordinated by one of us (AJCT). So far, successful 
photometric observations have been carried out for 4 GRB afterglows from 
UPSO, Nainital. The UPSO photometric observations for earlier 3 events namely
GRB 990123, GRB 991208 and GRB 991216 have been presented by Sagar et al. (1999,
2000). Such observations for the GRB000301C are presented here. It is worth
mentioning here that first earliest optical observations of GRB 000301C have 
been carried out by us. 
An introduction to the GRB 000301C studied here is given below.

\medskip
Smith et al. (2000) reported All Sky Monitor (ASM) on the  Rossi $X-$ray 
Timing Explorer (RXTE), Ulysses and Near Earth Asteroid Rendezvous (NEAR) 
detection of a GRB on 2000 March 01 at 09:51:37 UT. GRB 000301C therefore
joins the group of GRB 991208 (see Hurley et al. 2000) whose positions were 
determined only by the Interplanetary Network Localization alone without 
Compton Gamma-Ray observatory BATSE or BeppoSAX observations within 1.5 day of 
the event. The positions were of such accuracy ($\sim$50 arcmin$^2$ in this 
case) that it led to the successful identification of radio, millimeter, 
optical and near-IR afterglows, and eventually to the measurement of its 
redshift. Fig. 1 shows the light curve of the prompt $\gamma-$ray emissions of
the GRB 000301C detected by NEAR in the energy band 100 -- 1000 KeV downloaded 
from the {\sc http://lheawww.gsfc.nasa.gov/}. The burst profile is dominated by 
only one strong peak with no spike type structures generally observed with GRB 
events. The three ASM energy channels showed the strongest response in the 
5 -- 12 KeV band, reaching a peak flux of 3.7$\pm$0.7 Crab in 1 s time bin. 
Jensen et al. (2000) derived a peak flux of 6.3$\times 10^{-7}$ erg cm$^{-2}$ 
in 0.25 s time bin in the 25 -- 100 KeV energy range and the hardness ratio 
$\frac{f100-300}{f50-100}=2.7\pm0.6$ for the burst. It has a sharp rise and a 
relatively slow decline. Duration (full width at half maximum) of the profile 
at trigger of the burst is only $\sim$ 2 s. The detection of the GRB 000301C OT
was reported first by Fynbo et al. (2000a) in $U, B, R$ and $I$ passbands at 
$\alpha_{2000}=16^h 20^m 18.^s6; \delta_{2000}=+29^{\circ} 26^{'} 36^{''}$ 
with an uncertainty of $\sim 1^{''}$. It was confirmed by Bernabei et al. (2000)
on a $R$ image and by Stecklum et al. (2000) on a $K^{'}$ (2.1 $\mu$m) image. 
Coincident at the optical position, Bertoldi (2000) detected a flux of 
1.9$\pm$0.3 mJy at 250 GHz (1.2 mm) on 2000 March 4.385 UT while Berger \& 
Frail (2000) detected a flux of $\sim$ 300 $\mu$Jy at 8.46 GHz on 2000 March 
5.67 UT. No obvious emission or absorption features are visible in the low 
resolution spectrum of the OT of GRB 000301C taken on 2000 March 4.41 UT in the 
wavelength range of 410 -- 800 nm by Eracleous et al. (2000). An ultra-violet 
spectrum of the OT taken on 2000 March 6 with the Hubble Space Telescope (HST) 
by Smette et al. (2000) indicates a redshift of $z = 1.95\pm 0.1$. It was 
precisely determined to a value of $z = 2.0335\pm0.0003$ by Castro et al. (2000)
using a moderately high resolution spectra taken with the Keck-II 10-m telescope
on 2000 March 4. This determination was also supported by the low resolution 
spectrum obtained by Feng et al. (2000) on 2000 March 3.47 UT. The value is not
too different from 2.0404$\pm$0.0008 determined by Jensen et al. (2000) using 
absorption features in the spectrum obtained with
very large telescope on 2000 March 5 and 6. Rhoads \& Fruchter (2000), Masetti 
et al. (2000) and Jensen et al. (2000) present the near-infrared (IR) and 
$U, B, V, R$ and $I$ observations while Berger et al. (2000) provide broad-band 
(1.4 to 350 GHz) radio and millimeter wave observations. These data indicate 
that short term achromatic flux variability is superposed on the overall 
steepening of the light curve. The cause of the short term variability is not 
understood. However, it makes difficult the determination of break-time while
fitting jet model in the light curve of the afterglow emission. Therefore,
time scales determined in the published studies range from $\sim 3.5 - 7.5$
days. Present observations in combination with data 
published in the literature are used to study flux decay at optical and near-IR
wavelengths and spectral index from ultra-violet to radio regions. These data
have been used to determine precise parameters of the light curve.
Details of present optical observations etc. are given in the next section 
while light curves and other results are presented in the remaining sections.

\begin{figure*}
\begin{center}\vspace*{-0.5cm}\hspace*{-1cm}
\psfig{file=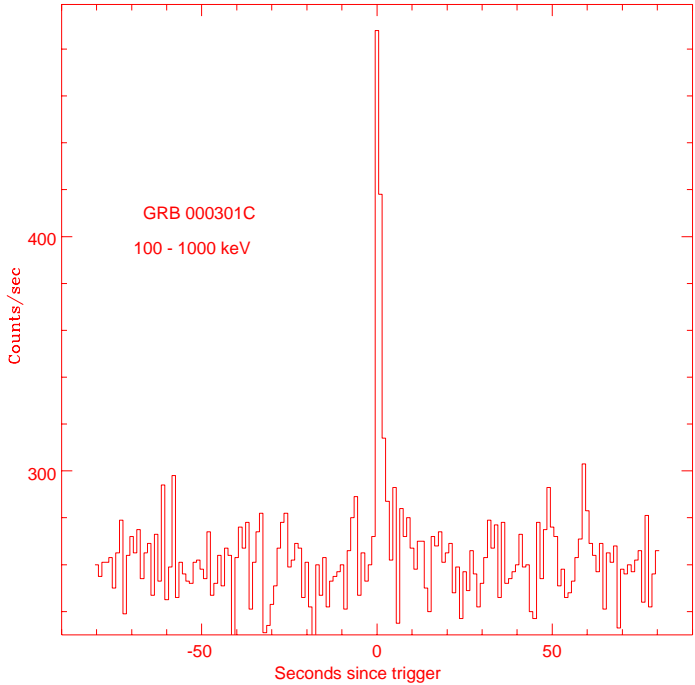,width=10cm,height=10cm}
\end{center}\vspace*{-0.6cm}
\caption{\label{prompt} 
The NEAR light curve of GRB 000301C in the energy range 100 -- 1000 keV. } 
\end{figure*}

\section { Optical observations, data reduction and calibrations } 

The optical observations of the GRB 000301C afterglow were carried out from 
2000 March 2 to 9. We used a 2048 $\times$ 2048 pixel$^{2}$ CCD system 
attached at the f/13 Cassegrain focus of the 104-cm Sampurnanand telescope of 
UPSO, Nainital. All the observations were done in good photometric sky 
condition, expect for 2000 March 6. One pixel of the CCD chip corresponds to 
0.$^{''}$38, and the entire chip covers a field of $\sim 13^{'} \times 13^{'}$ 
on the sky. Fig. 2 shows the location of the GRB 000301C afterglow on the CCD 
image taken from UPSO, Nainital. For comparison, image extracted from the 
Digital Palomar Observatory Sky Survey (DSS) is also shown where the absence 
of a GRB OT is clearly seen. 

\medskip
Several short exposures up to a maximum of 15 minutes were generally given. In 
order to improve the signal-to-noise ratio of the OT, the data have 
been binned in $2 \times 2$ pixel$^2$ and also all images of a night have been 
stacked after correcting them for bias, non-uniformity in the pixels and cosmic 
ray events. Exposure times for the stacked images were 70, 50, 85, 35, 105 and 
75 minutes in $R$ on 2000 March 2, 3, 5, 6, 7 and 8 respectively. Only one image
in each $V$ and $I$ filters could be taken on 2000 March 3 with corresponding 
exposure times of 30 and 10 minutes respectively. As the OT is close to 
a bright star, DAOPHOT profile-fitting technique is used for the magnitude 
determination. In the field of GRB 000301C, stars (as identified in Fig. 2) are 
photometrically calibrated in $R$ passband by Garnavich et al. (2000a). The 
quoted uncertainty in the zero-point calibration is $\pm$0.05 mag. Henden (2000)
provides the $UBVRI$ photometry for stars fainter than $R =$ 20 mag in the GRB 
000301C field. The $R$ magnitudes determined by Garnavich et al. (2000a) agree 
with an independent measurement reported by Henden (2000). This indicates that 
photometric calibration used in this work is secure. Present photometric 
magnitudes are relative to comparison star A and D (see Fig. 2). These along 
with other photometric measurements of GRB 000301C afterglow published by the 
time of paper submission are given in Table 1. In order to avoid errors arising 
due to different photometric calibrations, we have used only those published 
photometric measurements whose magnitudes could be determined relative to 
determinations given by either Garnavich et al. (2000a) or Henden (2000). 
In $JHK$ filters, all published photometric measurements have been used.

\begin{figure}
\begin{center}\vspace*{-0.8cm}\hspace*{-3cm}
\psfig{file=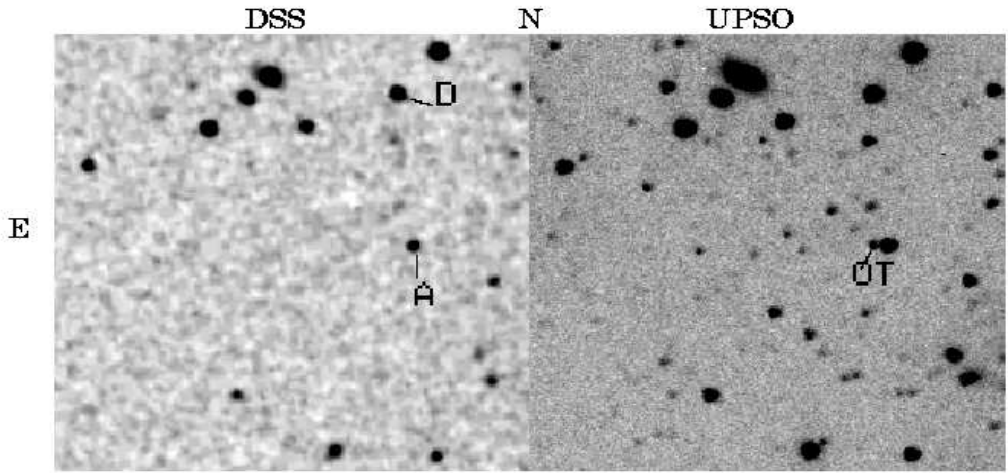,width=15cm,height=10cm}
\end{center}\vspace*{-2.0cm}
\caption{\label{ident} 
Finding chart for GRB 000301C field is produced from the CCD images taken from 
UPSO, Nainital on 2000 March 2.0 UT in $R$ filter with exposure time
 70 minutes. North is top and East is left.  The optical transient (OT) is 
marked on the UPSO image. Here only 1.$^{'}6 \times 1.^{'}$6 field of view is 
presented. The region corresponding to CCD image is extracted from the Digital 
Palomar Observatory Sky Survey and marked as DSS. A comparison of both images 
shows the absence of GRB afterglow on the DSS image. The comparison stars A 
and D (Garnavich et al. 2000a) are marked on the DSS image. } 
\end{figure}

\medskip
Present $R$ images have also been independently processed, reduced and 
calibrated by Masetti et al. (2000). A comparison of their $R$ values with ours
indicates good agreement. A small difference observed between the two sets of 
values is, perhaps, due to different data processing and calibration procedures.

\begin{table}
 {\bf Table 1.}~Photometric magnitues of the GRB 000301C afterglow. 

\begin{center}
\begin{tabular}{ccll} \hline 
Time in UT &Passband & Magnitude & Source  \\  \hline 
Mar 00 02.96&      B &      20.99$\pm$0.20&      Bhargavi \& Cowsik (2000) \\
Mar 00 03.18&      B &      21.07$\pm$0.05&  Masetti et al. (2000)\\
Mar 00 03.23&      B &      21.10$\pm$0.12&  Masetti et al. (2000)\\
Mar 00 03.50  &    B &      21.11$\pm$0.04&       Veillet  (2000a)  \\ 
Mar 00 04.52  &    B &      21.41$\pm$0.04&       Halpern et al. (2000c)  \\ 
Mar 00 04.91&      B &      21.31$\pm$0.14&      Bhargavi \& Cowsik (2000) \\
Mar 00 05.15&      B &      21.60$\pm$0.20&  Masetti et al. (2000)\\
Mar 00 06.16&      B &      22.45$\pm$0.15&  Masetti et al. (2000)\\
Mar 00 07.15&      B &      22.43$\pm$0.15&  Masetti et al. (2000)\\
Mar 00 14.60 &    B &      24.83$\pm$0.12&       Veillet  (2000d) \\ 
Mar 00 03.22&      V &      20.57$\pm$0.05&  Masetti et al. (2000)\\     
Mar 00 03.89  &    V &      20.95$\pm$0.06&       Present work \\
Mar 00 04.11  &    V &      21.10$\pm$0.06&       Gal-Yam et al. (2000)\\  
Mar 00 05.17&      V &      21.04$\pm$0.20&  Masetti et al. (2000)\\     
Mar 00 06.22&      V &      21.90$\pm$0.15&    Fruchter et al. (2000a)\\ 
Mar 00 02.93&      R &      20.42$\pm$0.06&      Present work \\
Mar 00 02.96&      R &      20.02$\pm$0.03&      Bhargavi \& Cowsik (2000) \\
Mar 00 03.14&      R &      20.25$\pm$0.05&  Masetti et al. (2000)\\
Mar 00 03.17&      R &      19.94$\pm$0.05&      Fynbo et al. (2000b)\\
Mar 00 03.19&      R &      20.16$\pm$0.05&  Masetti et al. (2000)\\    
Mar 00 03.21&      R &      20.25$\pm$0.05&  Masetti et al. (2000)\\   
Mar 00 03.51&      R &      20.24$\pm$0.05&        Halpern et al. (2000a)  \\ 
Mar 00 03.51&      R &      20.27$\pm$0.04&       Veillet  (2000a)  \\ 
Mar 00 03.51&      R &      20.28$\pm$0.05& Garnavich et al. (2000a)\\
Mar 00 03.93&      R &      20.53$\pm$0.06&            Present work \\
Mar 00 04.00&      R &      20.49$\pm$0.10&      Bhargavi \& Cowsik (2000) \\
Mar 00 04.04&      R &      20.53$\pm$0.06&   Masetti et al. (2000)\\     
Mar 00 04.08&      R &      20.57$\pm$0.06&      Gal-Yam et al. (2000)\\ 
Mar 00 04.38&      R &      20.56$\pm$0.05&       Garnavich et al. (2000b)\\ 
Mar 00 04.46&      R &      20.54$\pm$0.06&       Mujica et al. (2000)\\ 
Mar 00 04.50&      R &      20.61$\pm$0.04&       Halpern et al.  (2000b) \\ 
Mar 00 04.91&      R &      20.58$\pm$0.05&      Bhargavi \& Cowsik (2000) \\
Mar 00 05.14&      R &      20.47$\pm$0.07&  Masetti et al. (2000)\\
Mar 00 05.63&      R &      20.86$\pm$0.04&       Veillet  (2000a)\\
Mar 00 05.96&      R &      21.18$\pm$0.07&           Present work \\
Mar 00 06.14&      R &      21.65$\pm$0.20&  Masetti et al. (2000)\\  
Mar 00 06.22&      R &      21.50$\pm$0.15&        Fruchter et al. (2000a)\\ 
Mar 00 06.98&      R &      $> 21.8$      &          Present work \\
Mar 00 07.13&      R &      21.63$\pm$0.15& Masetti et al. (2000)\\
Mar 00 07.65&      R &      21.70$\pm$0.07&       Veillet  (2000b)\\ 
Mar 00 07.93&      R &      21.95$\pm$0.10&         Present work \\
Mar 00 08.15&      R &      21.68$\pm$0.10&   Masetti et al. (2000)\\ 
Mar 00 08.95&      R &      22.13$\pm$0.10&           Present work \\
Mar 00 09.52&      R &      22.28$\pm$0.09&       Halpern \& Kemp (2000)  \\ 
Mar 00 11.63&     R &      23.02$\pm$0.10&       Veillet  (2000c)\\  
Mar 00 14.60 &    R &      23.82$\pm$0.10&       Veillet   (2000d) \\  
Apr 00 03.90 & R    &      26.90$\pm$0.15 &  Fruchter et al. (2000b)\\ 
Mar 00 03.21&      I &      19.94$\pm$0.07&  Masetti et al. (2000)\\     
Mar 00 03.96&    I   &      19.94$\pm$0.15&       Present work \\
Mar 00 06.19&      I &      20.82$\pm$0.15&  Masetti et al. (2000)\\
Mar 00 07.18&      I &      21.20$\pm$0.15&  Masetti et al. (2000)\\   
Mar 00 08.17&      I &      21.61$\pm$0.10&  Masetti et al. (2000)\\ 
Mar 00 03.55&    J   &      18.88$\pm$0.02&       Kobayashi et al. (2000) \\
Mar 00 04.65&    J   &      19.06$\pm$0.05&       Rhoads \& Fruchter (2000) \\
Mar 00 03.22&    K$^{'}$   &17.51$\pm$0.06&       Rhoads \& Fruchter (2000) \\
Mar 00 03.56&    K$^{'}$   &17.52$\pm$0.02&       Kobayashi et al. (2000) \\
Mar 00 04.64& K$^{'}$ &   17.65$\pm$0.04&       Rhoads \& Fruchter (2000) \\
Mar 00 05.61& K$^{'}$ &   18.00$\pm$0.07&        Rhoads \& Fruchter (2000) \\
Mar 00 06.60& K$^{'}$ &   18.56$\pm$0.12&       Rhoads \& Fruchter (2000) \\
Mar 00 08.59& K$^{'}$ &   19.28$\pm$0.09&       Rhoads \& Fruchter (2000) \\
\hline
\end{tabular}
\end{center}
\end{table}

\section{ Optical and near-IR photometric light curves}

We have used the published data in combination with the present measurements to 
study the flux decay of GRB 000301C afterglow. Fig. 3 shows the plot of 
photometric measurements as a function of time. The X-axis is log ($t-t_0$) 
where $t$ is the time of observation and $t_0$ is the time of GRB burst which 
is 2000 March 1.411 UT. All times are measured in unit of day. Before deriving 
the flux decay constants of the OT, it is mandatory to subtract the 
contributions from foreground/background galaxies, if there is any. 
Both ground based  and the early HST images clearly show that any underlying 
galaxy would have to be fainter than $R >$ 25 mag (Fruchter et al. 2000a). 
In fact, the late-time HST images taken on 2000 April 3.9 UT by Fruchter et al. 
(2000b) showed that the $R$ magnitude of the host galaxy is 27.8$\pm$0.25. We 
have therefore not applied any correction upto $R <$ 23 mag of the OT for the 
contamination by host galaxy.

\begin{figure*}
\begin{center}\vspace*{-1.5cm}\hspace*{-1cm}
\psfig{file=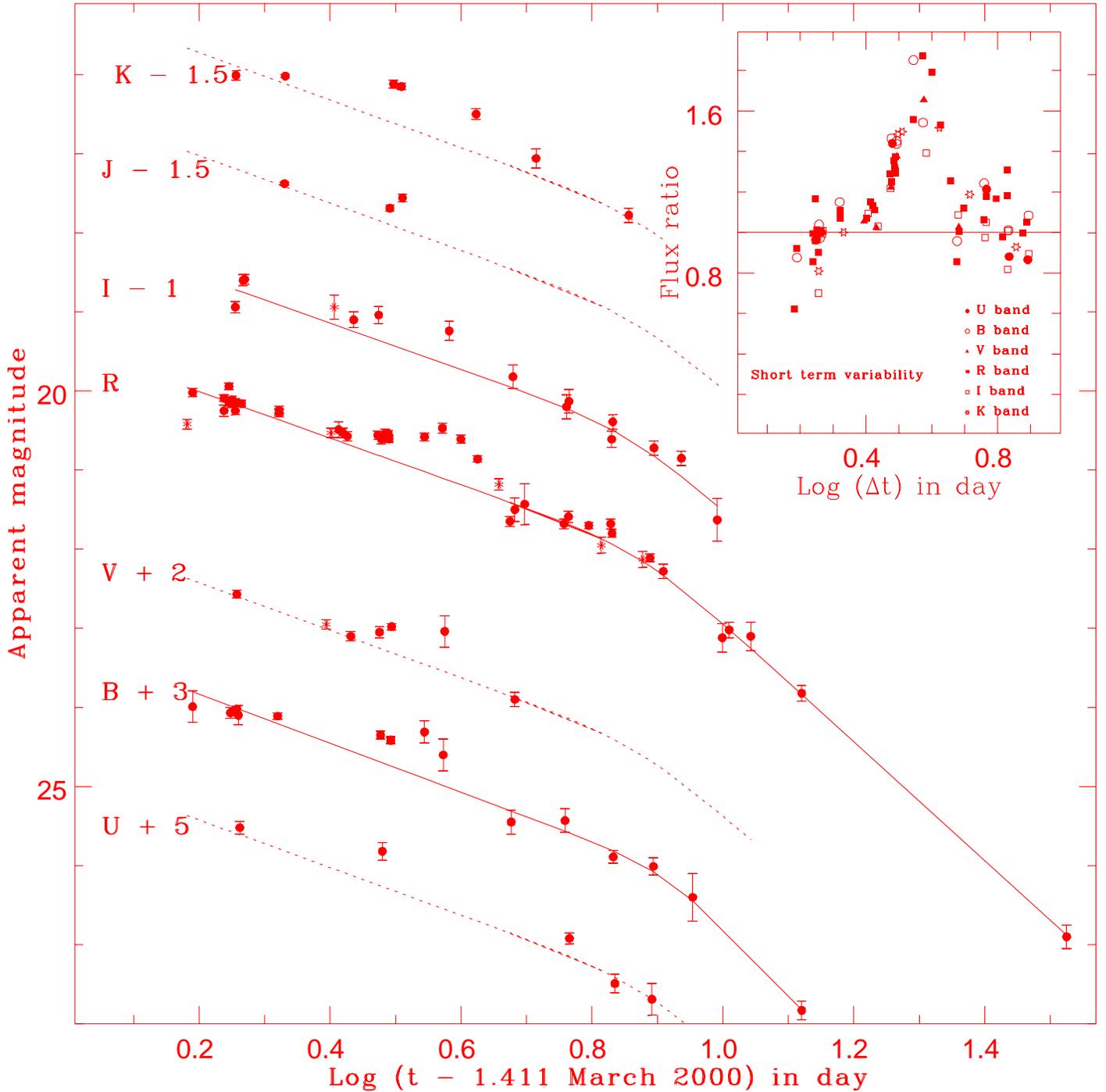,width=20cm,height=20cm}
\end{center}\vspace*{-1.1cm}
\caption{\label{light} Light curve of GRB 0000301C afterglow in optical and 
near-IR photometric passbands. Measurements from UPSO, Nainital have been 
indicated as asterisk. Suitable offsets have been applied to aviod overlapping 
in data points of different passbands. Flux decay can not be fitted by a single power-law. Solid line represents the least square non-linear fit to the 
densely observed data for a jet model while dotted lines are the jet model 
curves for $\alpha_1 = 1.2$ and $\alpha_2 = 3.0$. In all cases the value of
$\beta$ is taken as 5. Short term variability observed in different passbands 
is shown in the upper right corner box.} 
\end{figure*}

\medskip
The flux decay of most of the earlier GRB afterglows is generally well 
characterized by a single power law $F(t) \propto (t-t_0)^{-\alpha}$, where 
$F(t)$ is the flux of the afterglow at time $t$ and $\alpha$ is the decay 
constant. However, optical and near-IR light curves of GRB 000301C (Fig. 3) 
show erratic behaviour with an overall flux decay. Observers therefore took 
relatively long time to accept it as an OT of the GRB 000301C. 

\medskip
UPSO observation in $R$ filter on 2000 March 2.93 UT is the earliest 
optical observations published so far. Bhargavi \& Cowsik (2000) 
measurements are just after us. Fig. 3 clearly indicates peculiar behaviour of 
the light curve and perhaps, even shows $\sim$ 0.5 mag brightening of the $R$ 
magnitude with $\alpha = -0.5\pm1.0$ during $\Delta t = 1.5 - 1.8$ day. 
This could be an indication of a rising phase similar to that seen in GRB 970228
(Guarnieri et al. 1997) and GRB 970508 (Castro-Tirado et al. 1998). Contrary to 
most of the earlier GRB afterglows, light curve of GRB 000301C can not be fitted
by a single power-law (see also Masetti et al. 2000, Rhoads \& Fruchter 2000,
Berger et al. 2000; Jensen et al. 2000). Overall the OT flux decay seems to 
have broken power-law as expected in GRB afterglows having jet-like 
relativistic ejecta (Sari et al. 1999; Rhoads 1999). This appears to be 
superimposed with some shorter time flux variability especially during 
$\Delta t < 8$ day. Among equally well monitored GRB afterglows, GRB 000301C
appears therefore peculiar. Correlated variability can be clearly noticed in
$B, R$ and $I$ passbands. The lack of such apparent correlation in the light
curves of other passbands is most probably due to non-strict simultaneity of
the data points. Broken power-law in these light curves can be empirically 
fitted by functions of the form (see Rhoads \& Fruchter 2000) 

\medskip
$F(t) = 2F_0/[(t/t_b)^{\alpha_1\beta} + (t/t_b)^{\alpha_2\beta}]^{1/\beta}$, 

\medskip
where $\alpha_1$ and $\alpha_2$ are asymptotic power-law slopes at early and 
late times with $\alpha_1 < \alpha_2$ and $\beta > 0$. $\beta$ controls the
sharpness of the break, with larger $\beta$ implying a sharper break. With
$\beta = 1$, this function becomes the same that Stanek et al. (1999) fit the
optical light curve of GRB 990510 afterglow. $F_0$ is the flux at the cross-over
time $t_b$. The function describes a light curve falling as $t^{-\alpha_1}$ at 
$t << t_b$ and $t^{-\alpha_2}$ at $t >> t_b$. The function can be written in 
magnitudes as

\medskip
$m=m_b + \frac{2.5}{\beta} [log_{10} \{ (t/t_b)^{\alpha_1\beta}
+(t/t_b)^{\alpha_2\beta}\} - log_{10}(2)]$, 

\medskip
where $m_b$ is the magnitude at time $t_b$. In jet 
models, an achromatic break in the light curve is expected when the jet makes 
the transition to sideways expansion after the relativistic Lorentz factor 
drops below the inverse of the opening angle of the initial beam. Slightly
later, the jet begins a lateral expansion which causes a further steepening of
the light curve. Before fitting jet model to the light curve to derive
accurate flux decay parameters of the afterglow, it is mandatory to deconvolve 
the short term variability component. Otherwise, it will confuse the 
determination of $t_b$. Perhaps it is the main reason for having a range of
$t_b$ values in the literature. The short term variability component of the
light curve is determined as described below.

\medskip
In order to minimize the effects of short term variability on the determination
of the parameters of the fireball model, Berger et al. (2000) fitted the entire
data set from radio to optical simultaneously and derive $t_b = 7.5\pm0.5$ days,
$\alpha_1 = 1.28$ for $t < t_b$ and $\alpha_2 = 2.70$ for $t > t_b$ as global 
parameters for the jet fireball model. The observed $U, B, V, R, I,$ and 
$K^{'}$ fluxes are divided by the values obtained from the jet model fit 
yields, as also is noticed by Berger et al. (2000) that the variability is 
simultaneous and of similar amplitude in all bands (see upper right corner box 
in Fig. 3). There is a sharp rise and decline centered on $\Delta t = 4$ day. 
Berger et al. (2000) also found similar variability at 250 GHz. All these 
indicate that variability is the result of a real physical process which 
produces simultaneously similar level of absolute variation over a large range 
in frequency. Berger et al. (2000) therefore explain this fluctuation in terms 
of non-uniform ambient density which varied by about a factor of 3.

\medskip
We use the densely covered observations in $B$, $R$ and $I$ to determine the
parameters of jet model using the above function. For this, the short term 
variability was deconvolved from the observed light curves.
It has been noticed that the minimum value of $\chi^2$ is achieved
for $\beta \ge 5$. This indicates that the observed break in the light curve
is sharp and is unlike the smooth break observed in the optical light curve
of GRB 990510 (cf. Stanek et al. 1999; Harrison et al. 1999). In order to avoid
a fairly wide range of model parameters for a comparable $\chi^2$ due to
degeneracy between $t_b, \alpha_2$ and $\beta$, we have used fixed value of 
$\beta = 5$ in our further analyses. The least square best fitted parameters
$t_b, m_b, \alpha_1,$ and $\alpha_2$ have values 7.51$\pm$0.63, 22.15$\pm$0.15, 
1.18$\pm$0.14 and 3.01$\pm$0.53 respectively in $R$. The corresponding values 
are 8.27$\pm$1.11, 23.20$\pm$0.24, 1.24$\pm$0.20 and 3.48$\pm$2.07 respectively 
in $B$ and 7.27$\pm$1.04, 21.64$\pm$0.26, 1.17$\pm$0.29 and 2.92$\pm$2.93 
respectively in $I$. This indicates that average values of $t_b, \alpha_1,$ and 
$\alpha_2$ are 7.6$\pm$0.5 day, $1.2\pm0.1$ and $3.0\pm0.5$ respectively. The 
light curves derived with these averaged parameters using the jet model are 
shown by dotted curves in the $U, V, J$ and $K^{'}$ passbands. This clearly 
indicates the presence of simultaneous short term variability in all passbands.
We therefore conclude in agreement with Masetti et al. (2000) and
Rhoads \& Fruchter (2000) that optical and near-IR flux decays of GRB 000301C 
afterglow are peculiar in comparison to other such well observed GRB afterglows.

\subsection{Spectral index of the GRB 000301C afterglow }

The flux distribution of the GRB 000301C afterglow has been studied using the
broadband photometric measurements listed in Table 1 along with the published 
radio, millimeter and ultra-violet observations. We used the reddening map 
provided by Schlegel, Finkbeiner \& Davis (1998) for estimating Galactic 
interstellar extinction towards the burst and found a small value of $E(B-V) = 
0.05$ mag. We used the standard Galactic extinction reddening curve given by 
Mathis (1990) in converting apparent magnitudes into fluxes and used the 
effective wavelengths and normalisations by Bessell (1979) for $B, V, R$ and $I$
and by Bessell \& Brett (1988) for $J$ and $K^{'}$. The fluxes thus derived are 
accurate to $\sim$ 10\%. Fig. 4 shows the spectrum of GRB 000301C afterglow from
optical to radio region. The fluxes closest to $\Delta t = 4.8$ day at 1.4 GHz, 
4.86 GHz, 8.46 GHz, 15 GHz, 22.5 GHz, 100 GHz, 250 GHz and 350 GHz are taken 
from Berger et al. (2000). It is observed that as the frequency decreases the
flux increases from optical to millimeter wavelengths and then it turns over. 
The spectrum thus can be described by a single power law in some frequency
interval as $F_{\nu} \propto  \nu^{\beta}$, where $F_{\nu}$ is the flux at
frequency $\nu$ and $\beta$ is the spectral index. In the optical to millimeter 
region, the value of $\beta $ is $-0.73\pm$0.06 at $\Delta t =$ 4.8 day. The
optical-near-IR spectrum has not changed significantly (see Table 2 and Fig. 4) 
and has average value around $-1.0$. This is in agreement with a single value 
of $\beta = -1.1$ derived from the low-resolution spectrum taken on 2000 March 
3.47 UT by Feng et al. (2000) in the wavelength range of 0.3 to 0.6 $\mu$m. The 
HST observations taken around $\Delta t$ = 33.5 day by Fruchter et al. (2000b) 
also indicate similar slope. All these, perhaps, indicate no change in the 
spectral slope of GRB 000301C at later times. There is thus no evidence for a
cooling break passing through the optical band on these time scales. This is 
unlike GRB 980326 ( Bloom et al. 1999) and GRB 970228 (Fruchter et al. 1999; 
Galama et al. 2000) where spectral index changed to a value of 
$\beta \sim -3.0$ after $\Delta t >$ 20 days. The spectral slope at radio to 
millimeter frequencies is generally expected to be $+1/3$ at these early times. 
However, the observed slope is much larger with a value of $+0.90 \pm 0.08$. 
The peak frequency seems to lie in millimeter region. This peak frequency is 
thus similar to that of GRB 970508 (cf. Galama et al. 1998) but different from 
that of GRB 990123 (Galama et al. 1999) where the peak is in radio region and 
that of GRB 971214 for which the peak is in optical/near-infrared waveband 
(Ramaprakash et al. 1998). From this, one may infer that the synchrotron peak 
frequency may span a large range in GRB afterglows.

\begin{figure*}
\begin{center}\vspace*{-0.5cm}\hspace*{-1cm}
\psfig{file=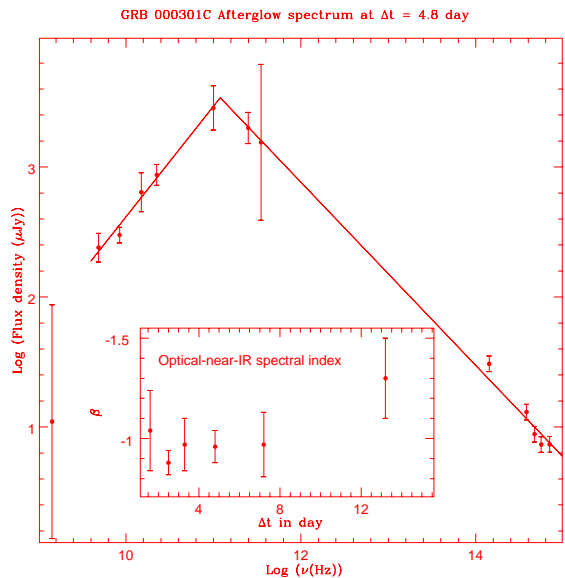,width=8cm,height=8cm}
\end{center}\vspace*{-0.6cm}
\caption{\label{spec} The spectral flux distribution of the GRB 000301C 
afterglow at $\sim$ 4.8 day after the burst. Fluxes measured at optical, 
near-IR, radio and millimeter wavelengths closest to the epoch are plotted. 
The least square linear relations derived using fluxes at optical, near-IR, 350 
GHz and 250 GHz and at 4.86 GHz, 8.46 GHz, 15 GHz, 22.5 GHz and 100 GHz are 
shown by solid lines. Inside the box, optical-near-IR spectral slopes derived 
at different epochs (see Table 2) are shown as a function of $\Delta t$.} 
\end{figure*}

\section{ Comparison with the synchrotron emisson model} 

It is generally believed that the observed afterglow results from slowing down 
of a relativistic shell on the external ISM and therefore is produced by 
external shocks. Recent afterglow observations of GRBs show that a 
relativistic blast wave, in which the highly relativistic electrons radiate
via synchrotron mechanism, provides a generally good description of the 
observed properties. Here we will discuss briefly the implications of the
observed flux decay exponent $\alpha$ and the spectral slope index $\beta$ in 
different wavelength range for such models.
All models have the flux as $F(\nu,t) \propto t^{-\alpha} \nu ^{\beta}$ for 
a range of frequencies and times that contain no spectral breaks. 
In each model $\alpha$ and $\beta$ are functions of $p$ only, the power-law 
exponent of the electron Lorentz factor. The measurement of either one of 
$\alpha$ or $\beta$ therefore fixes $p$ and predicts the other one.

\medskip
In order to study the expected changes in the spectral indices with $\alpha$,
we derive the value of $\beta$ in optical to near-IR region at different epochs.
They are plotted inside a box in Fig. 4 and listed in Table 2. Where necessary, 
flux measurements were interpolated between adjacent data points at one 
wavelength in order to determine a contemporaneous flux with another 
wavelength using the measurements listed in Table 1. There is no evidence 
for statistically significant large ($\Delta \beta \ge 0.5$) variation in the
spectral index on these time scales.

\begin{table}
 {\bf Table 2.}~Spectral slopes of GRB 000301C afterglow at selected epochs, 
$\Delta t$ in optical-near-IR region. Filters used in deriving the value of 
$\beta$ are listed. The values of $\alpha$ and $p$ predicted for spherical
 model are also listed.

\begin{center}
\begin{tabular}{clc cc} \hline 
Epoch &Filters & Observed  & \multicolumn {2}{c} { Predicted} \\ 
($\Delta t$ in days)& &$\beta\pm\sigma$&$\alpha\pm\sigma$&$p\pm\sigma$\\ \hline 
~1.6 & B, R & $-1.04\pm$0.20  & 1.56$\pm$0.30 & 3.08$\pm$0.40 \\
~2.5&B, V, R, I, J, K$^{'}$&$-0.88\pm$0.06 &1.32$\pm$0.09 & 2.76$\pm$0.12  \\
~3.3 & B, V, R,I, J, K$^{'}$ & $-0.97\pm$0.13 &1.46$\pm$0.20 & 2.94$\pm$0.26  \\
~4.8 & B, V, R, I, K$^{'}$ & $-0.96\pm$0.08 & 1.44$\pm$0.12 & 2.92$\pm$0.16  \\
~7.2 & B, R, K$^{'}$ & $-0.97\pm$0.14 & 1.48$\pm$0.21 & 2.94$\pm$0.28  \\
13.2 & B, R & $-1.30\pm$0.20  & 1.95$\pm$0.30 & 3.60$\pm$0.40 \\ \hline
\end{tabular}
\end{center}
\end{table}

\medskip
For comparison with model predictions, we assume that our observations are in 
the slow cooling regime and the $\nu_m$ has passed optical but not the cooling 
frequency, $\nu_c$ which most probably lies above optical region. Following
Sari et al. (1999), values of $\alpha$ and $p$ are predicted using observed
value of $\beta$ for the spherical model of the afterglow. They are listed in 
Table 2. The observed flux decay constant at early times agrees well with the
predicted ones given in Table 2 while exactly 
opposite is the case at late times for spherical afterglow emission. But
the value of flux decay constant $\alpha$ is expected to approach the 
electron energy distribution index $p$, when the evolution of the afterglow is 
dominated by the spreading of the jet. On the other hand, the value of $\beta$
is the same for both spherical and jet models. Since the observed values of
$\alpha$ for late times agree with the predicted values of $p$ and hence
to the values of $\alpha$ in jet model, we conclude that afterglow emission 
from GRB 000301C is of jet type and not spherical.

\section{ The energetics of the GRB 000301C}

Redshift determination of $z = 2.0335\pm0.0003$ (Castro et al. 2000) for the 
GRB 000301C afterglow yields a minimum luminosity distances  of 16.6 Gpc for a 
standard Friedmann cosmological model with Hubble constant $H_0$ = 65 km/s/Mpc, 
cosmological density parameter $\Omega_0$ = 0.2 and cosmological constant 
$\Lambda_0$ = 0 (if $\Lambda_0 > 0$ then the inferred distances would increase).
The GRB 000301C thus becomes the second farthest GRB after GRB 971214 (Kulkarni 
et al. 1998) amongst the GRBs with known redshift measurements so far.

\medskip
As there is no published observed fluence in any energy range for this GRB, we 
estimate it indirectly assuming that present GRB event may also have 
the ratio between optical flux density and gamma-ray energy fluence similar to
those observed so far which is $\ge 10^{-23}$ (see Table 3 in Briggs et al. 
1999). Taking $R = 20$ mag at $\Delta t \sim$ 1 day, this ratio yields an 
energy fluence of at least $10^{-5}$ ergs/cm$^2$ above 20 KeV. Considering 
isotropic energy emission and this observed fluence and using the inferred 
luminosity distances, we estimate the $\gamma-$ray energy release to be at least
$3.4 \times 10^{53}$ ergs $\sim 0.2 M_{\circ}c^2$ for this GRB. Considering
the different fluence energy ranges used, this is not too
different from the values $\sim 5.4 \times 10^{52}$ ergs and $2.27 \times 
10^{52}$ ergs derived by Breger et al. (2000) and Jensen et al. (2000) 
respectively. Theoretical models predict that brightness of the prompt optical 
flash can be as bright as 9 -- 10 mag (Sari \& Piran 1999); as was observed in 
the case of the GRB 990123, the only prompt optical emission detected so far. 
At the optical distance of GRB 000301C, this implies a peak optical luminosity 
of $\sim 6.3 \times 10^{16}$ times the solar luminosity, if the prompt optical 
emission is of similar order. This is about a million times the luminosity of 
a normal galaxy and about a thousand times the luminosity of the brightest 
quasars known. The present energy and $t_b$ estimates imply a jet opening 
angle of $0.15 n^{1/8}$ radian, where $n$ is the number density (in 
units of cm$^{-3}$) of the ambient medium. This means that the actual energy 
released from the GRB 000301C is reduced by a factor of $\sim 90$ relative to 
the isotropic value and becomes $\sim 3.8 \times 10^{51}$ ergs.

\medskip
Of the over dozen GRBs with known redshifts, six with total fluence energies 
$>$ 20 keV in excess of 10$^{53}$ erg (assuming isotropic emission) are 
GRB 000301C (discussed here); GRB 991216 and GRB 991208 (Sagar et al. 2000); 
GRB 990510 (Harrison et al. 1999); GRB 990123 (Andersen et al. 1999; Galama et 
al. 1999) and  GRB 971214 (Kulkarni et al. 1998). Recent observations suggest 
that GRBs are associated with stellar deaths, and not with quasars or the 
nuclei of galaxies as some GRBs are found off-set from their host galaxy. 
However, release of huge amount of isotropic energy of $\sim 10^{53}$ erg or 
more is essentially incompatible with the popular stellar death models
(coalescence of neutron stars and death of massive stars). Recent observations 
seem to indicate non-isotropic emission as the most plausible way to reduce the 
enormous energy release. Indeed, almost all energetic 
sources in astrophysics such as pulsars, quasars and accreting stellar black 
holes display jet-like geometry and hence, non-isotropic emission. Beaming 
reduces the estimated energy by a factor of 10 - 300, depending upon the size 
of its opening angle (Sari et al. 1999). The $\gamma-$ray energy released then
becomes $\leq 10^{52}$ erg, a value within reach of current popular models for 
the origin of GRBs (see Piran 1999 and references therein).

\section{Discussions  and Conclusions }

 Prompt $\gamma-$ray emission light curve of the GRB 000301C burst shows, 
unlike most of the GRB events, only one strong peak with a flux of 3.7$\pm$0.7 
Crab in the 5 -- 12 KeV energy range. Using optical and near-IR observations, 
we obtained the values of flux decay constants and spectral indices. Light 
curves of the GRB 000301C afterglow emissions are peculiar. The light curves 
show a steepening superposed by a short term flare type variability which 
could be detected mainly due to the dense observations in $R$ filter.
A large fraction of these observations have been carried out using the 1-m 
class optical telescopes. This indicates that in future these telescopes, as 
large amount of observing time is available on them, will  play an important
role in understanding the origin of such short term variability in the light
curves of GRBs during early times. The overall flux decay in observed light 
curves are well understood in terms of a jet model. The parameters of the jet 
model are derived by fitting least square non-linear fit to the light curves 
obtained after deconvolving the short term variability from the observed light 
curves. The flux decay constants at early and late times are 1.2$\pm$0.1 and 
3.0$\pm$0.5 respectively. The value of jet break time is 7.6$\pm$0.5 day.
Before deriving any further conclusions from the light 
curve of GRB 000301C afterglow, we compare it with other well studied GRBs. 
Except GRB 990123, GRB 990510 and GRB 991216, all exhibit, at both early and 
late times a single power-law decay, generally $\sim$ 1.2, a value reasonable 
for spherical expansion in the fireball synchrotron model. GRB 000301C 
thus becomes the fourth burst for which a strong break in the light 
curve is clearly observed. Such breaks were observed first in the optical light 
curves of the afterglow of GRB 990123 (Castro-Tirado et al. 1999; Kulkarni et 
al. 1999) and recently in that of GRB 990510 (Harrison et al. 1999, Stanek et 
al. 1999) and GRB 991216 (Halpern et al. 2000, Sagar et al. 2000). They have 
generally been considered as evidence for collimation of the jet-like 
relativistic GRB ejecta in accordance with the prediction by recent theoretical 
models (M\'{e}sz\'{a}ros \& Rees 1999; Rhoads 1999; Sari et al. 1999). 

\medskip
The quasi-simultaneous spectral energy distributions determined in optical and 
near-IR regions for various epochs indicate that spectral index of the GRB 
000301C afterglow has not changed significantly  during a period of about 
35 days after the burst. The value of $\beta$ is $\sim -1.0$. However, the early
time flux decay constant has varied from 1.2$\pm$0.1 to 3.0$\pm$0.5. A 
steepening of flux decay constant with no corresponding change in spectral
index is attributed to the presence of a jet in the GRB 000301C OT. The jet
breaks around 7.6 days after the burst.

\medskip
Redshift determination yields a minimum distance of 16.6 Gpc, if one assumes 
standard Friedmann cosmology with $H_o = 65$ km/s/Mpc, $\Omega_0 = 0.2$ and 
$\Lambda_0 = 0$. GRB 000301C is thus at cosmological distance and becomes the 
second farthest amongst the GRBs with known distances so far. Considering 
isotropic energy emission, we estimate enormous amount of the $\gamma-$ray 
energy release ($\ge 10^{53}$ erg) above 20 KeV. This high energy is reduced
to $< 10^{52}$ erg when effects of non-isotropic emission are considered due
to the presence of a jet of an opening angle of 0.15 radian in the GRB 000301C.

\medskip
The peculiarity in the light curves of GRB 000301C seems to be due to 
superposition of a short term achromatic variability over a large frequency 
range on the overall steepening in the flux of the GRB 000301C. In separating 
the two components of the observed
light curves, dense as well as multi-wavelength observations during early 
times have played major role. Such observations of recent GRBs have started
revealing features which require explanations other than generally accepted 
so far indicating that there may be yet new surprises in GRB afterglows. 

\bigskip
\noindent {\bf Acknowledgements:} 
Suggestions/comments given by anonymous referee improved the scientific
content of the paper significantly. This research has made use of data 
obtained through the High Energy Astrophysics Science Archive Research Center 
Online Service, provided by the NASA/Goddard Space Flight Center.

\medskip
\noindent {\bf References:} 
\begin{itemize}
\item []Andersen M. I.  et al., 1999, Science, {\bf 283}, 2075
\item [] Bernabei S. et al., 2000, GCN Observational Report No.	571
\item [] Bertoldi F., 2000, GCN Observational Report 580
\item [] Bessell M.S., 1979, PASP, {\bf 91}, 589
\item [] Bessell M.S., Brett J.M., 1988, PASP, {\bf 100}, 1134
\item [] Bhargavi S.G., Cowsik, R., 2000, GCN Observational Report No. 630
\item [] Bloom J.S. et al., 1999, Nature, {\bf 401}, 453
\item [] Berger E., Frail, D.A., 2000, GCN Observational Report No. 589
\item [] Berger E. et al., 2000, Astro-ph/0005465
\item [] Briggs M.S. et al., 1999,  ApJ, {\bf 524}, 82
\item[] Castro S. M. et al., 2000, GCN Observational Report No. 605
\item [] Castro-Tirado A. J.  et al., 1998, Science, {\bf 279}, 1012
\item [] Castro-Tirado A. J.  et al., 1999, Science, {\bf 283}, 2069
\item [] Eracleous, M., Shetrone M., Sigurdsson S., Meszaros P., Wheeler J.C., 
Wang L., 2000, GCN Observational Report 584 
\item[] Feng M. et al., 2000, GCN Observational Report No. 607 
\item [] Fruchter A.S. et al., 1999, ApJ, {\bf 516},  683
\item [] Fruchter A.S. et al., 2000a, GCN Observational Report No. 602 
\item [] Fruchter A.S. et al., 2000b, GCN Observational Report No. 627 
\item [] Fynbo J.P.U., Jensen B.L., Hjorth J., Pedersen H., Gorosabel J., 2000a,      GCN Observational Report No. 570
\item [] Fynbo J.P.U., Jensen B.L., Hjorth J., Pedersen H., Gorosabel J., 2000b,      GCN Observational Report No. 576
\item [] Galama, T.J. et al., 1998, ApJ, {\bf 500}, L97
\item [] Galama T.J. et al., 1999, Nature, {\bf 398}, 394
\item [] Galama T.J. et al., 2000, ApJ Submitted
\item [] Gal-Yam A., Ofek E., Maoz D., Leibowitz, E.M., 2000, GCN Observational 
  Report No. 593
\item [] Garnavich P., Barmby P., Jha S., Stanek K., 2000a, GCN Observational 
  Report No. 573
\item [] Garnavich P., Barmby P., Jha S., Stanek K., 2000b, GCN Observational 
  Report No. 581
\item [] Guarnieri et al., 1997, A\&A, {\bf 328}, L13
\item [] Halpern J.P., Mirabal N., Lawrence, S., 2000a,	GCN Observational Report No. 578
\item [] Halpern J.P., Mirabal N., Lawrence, S., 2000b,	GCN Observational Report No. 582
\item [] Halpern J.P., Mirabal N., Lawrence, S., 2000c,	GCN Observational Report No. 585
\item[]Halpern J.P., Kemp J., 2000,  GCN Observational Report No. 604 
\item[]Halpern J.P. et al., 2000, ApJ Submitted
\item[]Harrison F.A. et al., 1999, ApJ, {\bf 523}, L121
\item [] Henden A., 2000, GCN Observational Report 583
\item [] Hurley K., et al., 2000, ApJ Lett (accepted) 
\item [] Jensen B.L. et al., 2000, A\&A Submitted/astro-ph/0005609
\item [] Kobayashi N., Goto M., Terada H., Tokunaga A.T., 2000, GCN Observational Report No. 577
\item [] Kulkarni S.R. et al., 1998, Nature, {\bf 393}, 35
\item [] Kulkarni S.R. et al., 1999, Nature, {\bf 398}, 389
\item [] Masetti N. et al., 2000, A\&A Submitted/astro-ph/0004186
\item [] Mathis J.S., 1990, ARAA, {\bf 28,} 37
\item [] M\'{e}sz\'{a}ros P., Rees M. J., 1999, MNRAS, {\bf 306}, L39
\item [] Mujica R., 2000, GCN Observational Report No. 597
\item [] Piran T., 1999, Physics Reports {\bf 314},  575 
\item [] Ramaprakash, A.N. et al., 1998, Nature, {\bf 393}, 43
\item [] Rhoads J.E., 1999, ApJ, {\bf 525}, 737
\item [] Rhoads J.E., Fruchter A.S., 2000, ApJ Submitted/astro-ph/0004057
\item [] Sagar R., 2000, Current Science {\bf 78}, 1076
\item [] Sagar R., Pandey A.K., Mohan V., Yadav R.K.S., Nilakshi, Bhattacharya 
D., Castro-Tirado A.J.,  1999, BASI, {\bf 27}, 3
\item [] Sagar R., Mohan V., Pandey A.K., Pandey S.B., Castro-Tirado A.J.,  2000, BASI, {\bf 28}, 15
\item [] Sari R., Piran T., 1999, ApJ, {\bf 517}, L109
\item [] Sari R., Piran T., Halpern J. P., 1999, ApJ, {\bf 519}, L17
\item [] Schlegel D.J., Finkbeiner D.P., Davis M., 1998, ApJ, {\bf 500}, 525
\item[]  Smette A. et al., 2000, GCN Observational Report No. 603
\item [] Smith D.A., Hurley K., Cline T., 2000, GCN observational Report No. 568
\item [] Stanek K. Z. et al., 1999, ApJ {\bf 522}, L39
\item[]  Stecklum B. et al., 2000, GCN Observational Report No. 572
\item [] Veillet C., 2000a, GCN Observational Report No. 588
\item [] Veillet C., 2000b, GCN Observational Report No. 598
\item [] Veillet C., 2000c, GCN Observational Report No. 610
\item [] Veillet C., 2000d, GCN Observational Report No. 611
\end{itemize}
\end{document}